\documentclass[aps,prl,twocolumn,groupedaddress]{revtex4}
\usepackage{graphicx}% Include figure files

\begin{document}

\title{Surface Acoustic Wave Propagation and Inhomogeneities in Low Density Two--Dimensional Electron Systems Near the Metal-Insulator Transition}

\author{L. A. Tracy$^1$, J.~P. Eisenstein$^1$, M.~P. Lilly$^2$, L.~N. Pfeiffer$^3$, and K.~W. West$^3$}

\affiliation{$^1$California Institute of Technology, Pasadena CA 91125 
\\
$^2$Sandia National Laboratories, Albuquerque, NM 87185	 
\\
$^3$Bell Laboratories, Lucent Technologies, Murray Hill, NJ 07974\\}

\date{\today}

\begin{abstract}
We have measured the surface acoustic wave velocity shift in a GaAs/AlGaAs heterostructure containing a two-dimensional electron system (2DES) in a low-density regime ($<$ $10^{10}$ cm$^{-2}$) at zero magnetic field.  The interaction of the surface acoustic wave with the 2DES is not well described by a simple model using low-frequency conductivity measurements.  We speculate that this conflict is a result of inhomogeneities in the 2DES which become very important at low density.  This has implications for the putative metal-insulator transition in two dimensions.  
\end{abstract}


\maketitle

The apparent metal-insulator transition\cite{krav} in two-dimensional electron systems (2DES) continues to attract intense interest, in large part because the basic physics is still not understood. For example, is the ``metallic'' phase  essentially a weakly localized Fermi liquid or an exotic non-Fermi liquid stabilized by strong Coulomb interactions\cite{abrahams,dassarma1}?  Is the transition itself governed by the straightforward physics of classical percolation of 2D electrons in a random potential\cite{efros0,nixon}, or is a more exotic zero temperature quantum phase transition at work\cite{abrahams}?  Although experimental evidence for inhomogeneities in the 2D density has begun to emerge\cite{yacoby}, and support the percolation picture\cite{meir,dassarma2}, the issue is not yet closed.

In order to address these questions we have examined the conductivity and homogeneity of low density 2D electron systems using two very different, and somewhat unconventional, probes.  First, the propagation of high frequency (0.1 -– 1 GHz) surface acoustic waves (SAW) across the region of the sample containing the 2D electron gas was observed.  As the 2D system is depleted, by increasing the (negative) voltage on a metal gate electrode on the heterostructure surface, the SAW experiences a velocity shift.  This velocity shift reflects both the local conductivity of the 2DES and, when the system is strongly inhomogeneous, the fraction $f$ of the sample area occupied by 2D electrons.  Second, the complex admittance $Y$ of the 2DES/gate system as a function of 2D density was measured at low frequencies (10 Hz –- 10 kHz). The gate capacitance and 2D sheet conductivity were extracted from such measurements by applying a distributed RC circuit model.  An advantage of both the SAW and capacitive techniques is their high sensitivity to very small conductivities. We find however, that the two techniques yield dramatically different results at very low density and argue that this conflict reflects the breakup of the 2D system into a landscape of puddles and dry spots.  In particular, we stress that the SAW technique, which is contactless, is sensitive to the presence of isolated patches of 2D electrons which develop when the average density of the system is reduced below the percolation threshold for bulk conduction.
 
\begin{figure}[b]
\includegraphics[width=3.5 in,bb= 71 80 341 269]{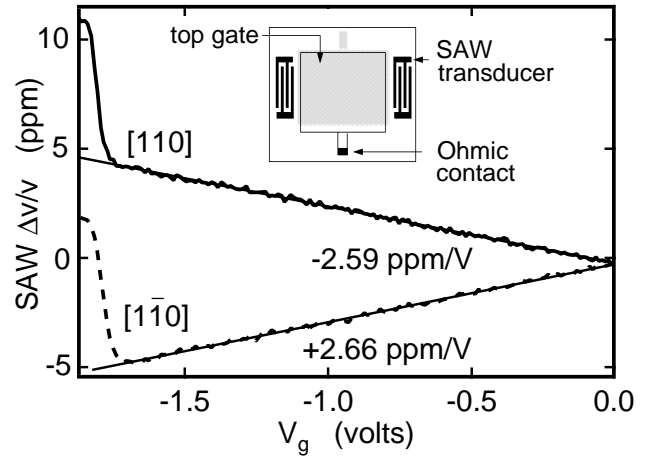}
\caption{SAW fractional velocity shift $\Delta v/v$ at 671 MHz and $T = 300$ mK as a function of gate voltage in two 2DES samples cut from the same MBE wafer. The two traces correspond to orthogonal propagation directions relative to the GaAs crystal axes.  Linear portions of the data for $V_g > -1.6$ V reflect a strain effect while upward jumps near -1.7 V signal the depletion of the 2DES.
 \label{fig1}}
\end{figure}
The samples used in this experiment are conventional modulation-doped GaAs/AlGaAs heterostructures grown by molecular beam epitaxy (MBE).  The great majority of the data presented here were obtained from a single-interface structure containing a 2DEG buried $d = 600$ nm beneath the sample surface.  In its as-grown state the 2DES has a sheet density of $N_s = 1.4 \times 10^{11}$ cm$^{-2}$ and a mobility of about $3 \times 10^6$  cm$^{2}$/Vs at low temperatures.  The 2DES is confined to a square mesa, 2 mm on a side. A single arm extends outward from the midpoint of one side of this mesa to a diffused In ohmic contact.  An aluminum gate electrode covers most of the square mesa, although a narrow strip along the side containing the contact arm is left uncovered.  When the gate is biased to reduce the electron density underneath it, this uncovered strip, which remains at the as-grown density, serves to establish an approximate equipotential along one edge of the gated region. This simplifies the distributed-circuit analysis of the gate-2DES admittance measurements. Two interdigitated SAW transducers are positioned off mesa, one on each side as shown in the inset of Figure \ref{fig1}.  Standard homoodyne techniques are used to measure the attenuation and velocity shift of SAWs propagated across the 2DES.  The fundamental frequency of the transducers is 120 MHz, but measurements up to $\sim$ 1.3 GHz were performed using transducer harmonics.  SAW and gate admittance measurements were done at excitation levels low enough to ensure linear response. Simultaneous SAW and gate admittance measurements were performed at $T$ = 300 mK in a $^3$He immersion cryostat. Additional admittance measurements (on the same sample) were also performed at temperatures down to 50 mK in a dilution refrigerator.

Due to the piezoelectricity of GaAs, SAWs interact with a 2DES near the sample surface, experiencing both attenuation and velocity shifts dependent upon the frequency and wavevector dependent conductivity $\sigma_{xx}(q,\omega)$ of the electron gas.  Previous SAW studies of 2DESs in the integer and fractional quantum Hall effect regime have shown that the SAW response is well-described by a homogeneous dc conductivity model\cite{Wixforth,Willett}.  In this model the expected relation between the fractional velocity change and the conductivity $\sigma$ of the 2DES is
\begin{eqnarray} 
\frac {\Delta v}{v} = \frac {\alpha^2/2}{1+(\sigma_{xx} /\sigma _{M})^{2}}
\label{eq:one}
\end{eqnarray}
where $\alpha$ is a constant reflecting the strength of the piezoelectricity in GaAs, the product $qd$ of the SAW wavevector $q$ and depth $d$ of the 2DES beneath the sample surface, and the presence of a highly conducting metal gate on that surface.  The characteristic conductivity $\sigma_M$, which reflects the ability of the 2DES to screen the SAW electric field, also depends on $qd$, but is typically $\sigma_{M}\approx 7 \times 10^{-7}~\Omega^{-1}$ under the conditions of the present experiments\cite{simon}.  The smallness of $\sigma_M$ implies that, at least at zero magnetic field, SAW velocity shifts will occur only when the 2DES is close to depletion.

Figure \ref{fig1} shows the measured SAW fractional velocity shift $\Delta v/v$ at 671 MHz as  a function of top gate voltage in two samples cut from a single $\langle 100 \rangle$-oriented GaAs wafer. The solid and dashed traces refer to SAW propagation along the $\langle 110 \rangle$ and $\langle 1 \bar{1} 0 \rangle$ crystallographic directions on the surfaces of these samples.  For gate voltages $V_g > -1.6$ V a simple linear dependence of $\Delta v/v$ is observed, with opposite signs for the two different crystal directions.  Unrelated to the conductivity of the 2DES, this effect reflects a gate-induced uniaxial strain in the GaAs sample: An electric field applied along $\langle 100 \rangle$ slightly contracts the sample along $\langle 110 \rangle$ while simultaneously expanding it along $\langle 1 \bar{1} 0 \rangle$.  The observed slopes of the data are consistent with simple modeling of the induced strain. We remark that this gate-induced rotational symmetry breaking may have implications for the anisotropic electronic phases recently discovered in high mobility 2DESs at high Landau level occupancy\cite{lilly}.  Figure 1 also shows that when $V_g$ is reduced below -1.6 V, both traces exhibit a jump upward in the velocity shift.  This jump signals the depletion of the 2DES under the gate.  While the data shown in Fig. 1 were obtained using a sample containing a double layer 2DES, the same qualitative effects are observed in the single layer 2D system, described earlier, to which we now turn.

\begin{figure}[t]
\includegraphics[width=3.5in, bb=134 90 413 284]{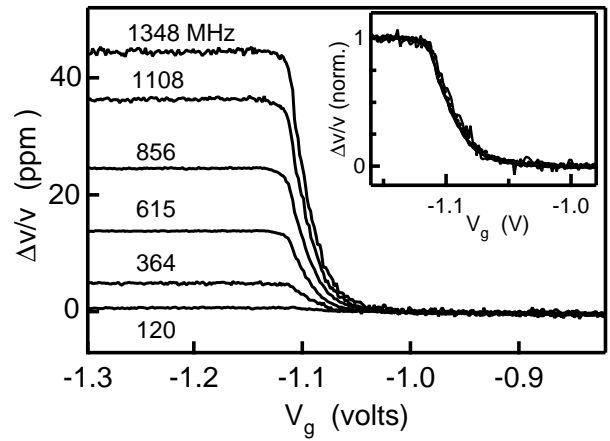}
 \caption{Depletion-induced jump SAW velocity shift in a single layer 2DES sample vs. gate voltage at various frequencies.  Inset:  Collapse of same data sets onto a single curve after normalization by the net jump in $\Delta v/v$ at each frequency.  The 120 MHz data is not included in the inset owing to its much lower signal-to-noise ratio.  
 \label{fig2}}
\end{figure}
The SAW velocity increases upon depletion of a buried 2DES because the piezoelectric field accompanying the elastic wave is no longer screened by the electron gas.  Since the SAW electric field penetrates to a depth of about one wavelength, the depletion-induced jump in $\Delta v/v$ is frequency dependent\cite{strain}.  Figure 2 shows this dependence in our single layer 2DES sample. Interestingly, the various curves all have the same shape; by normalizing each jump to unit height, all the data collapse onto a single curve.  We shall return to this point later.

\begin{figure}[b]
\includegraphics[width=3.5in, bb=0 0 270 152]{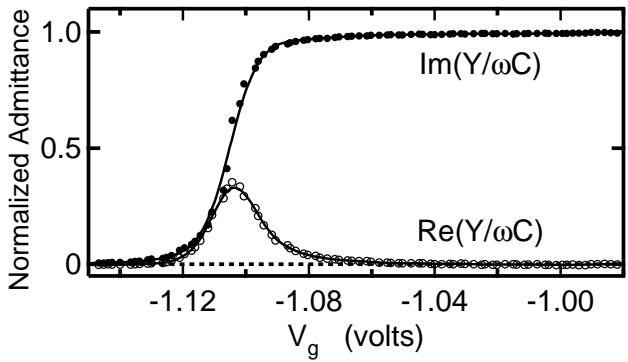}
 \caption{Real and imaginary parts of the gate admittance, $Y$, normalized by $\omega C$, at 1.3 kHz. The low frequency conductivity $\sigma_{xx}$ is extracted from such data near the peak in Re($Y/\omega C$).
 \label{fig3}}
\end{figure}
Figure 3 displays gate admittance data taken at $T$ = 300 mK with 1.3 kHz excitation.  The real and imaginary parts of these typical data are plotted versus the dc gate voltage $V_g$.  Both traces have been normalized by $\omega C$, the product of the (angular) measurement frequency and the gate capacitance $C = 670$ pF, itself determined from the admittance near $V_g = 0$ where resistive effects are negligible.  As depletion is approached Im($Y$) falls while Re($Y$) exhibits a maximum.  This behavior reflects the increasing sheet resistance $R$ of the 2DES, and the peak in the real part of the admittance occurs where $\omega RC \approx 1$.  In order to quantitatively determine the sheet conductivity $\sigma_{xx}$ from data such as these, we model the device as a 1-dimensional distributed RC circuit.  While such modelling in principle allows determination of the full $V_g$ dependence of $\sigma_{xx}$ from admittance data taken at a single frequency $\omega$, we instead only extract values of $\sigma_{xx}$ at the gate voltages $V_g$ surrounding the peak in Re($Y$). The frequency is then changed, $Y(V_g)$ re-measured, and the analysis repeated.  In this way a complete determination of $\sigma_{xx}(V_g)$ is obtained. 

Figure 4 shows $\sigma_{xx}$, in units of the conductance quantum $e^2/h \approx 3.9 \times 10^{-5}~\Omega^{-1}$, versus $V_g$ and 2D density $N_s$ at $T = 400$, 150, and 50 mK.  The calibration of density versus gate voltage was obtained by observing the quantum oscillations in Im($Y$) induced by a small magnetic field applied perpendicular to the 2DES\cite{calib}. The figure shows that for densities $N_s > 8 \times 10^9 \rm cm^{-2}$, the 8-fold change in temperature has negligible effect on the conductivity, suggesting that the 2DES is in the ``metallic'' phase.  At lower densities however, the conductivity falls with temperature, ever more rapidly as $V_g$ and $N_s$ are reduced.  For $N_s < 6 \times 10^9 \rm cm^{-2}$ the 2DES in our sample is almost certainly an insulator in the $T = 0$ limit.  These data suggest that the metal-insulator transition in our sample occurs at $N_s \approx 7 \times 10^9 \rm cm^{-2}$ where, interestingly, $\sigma_{xx} \approx e^2/h$.

\begin{figure}[t]
\includegraphics[width=3.5in, bb=0 0 281 180]{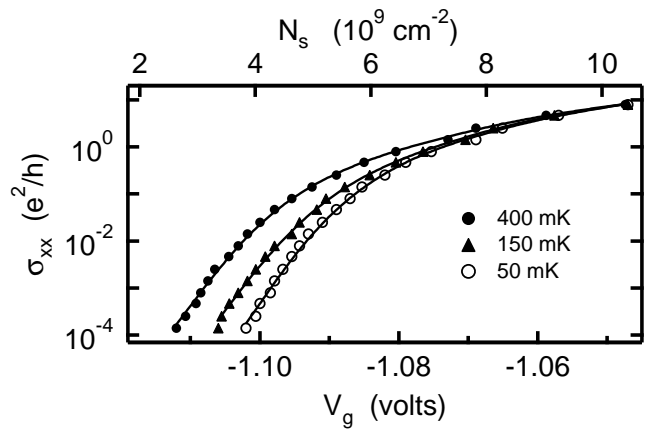}
 \caption{Conductivity vs. gate voltage (and density) at three temperatures.  For densities below about $7 \times 10^9 \rm cm^{-2}$ the 2DES appears to have an insulating ground state.
 \label{fig4}}
\end{figure}
Surface acoustic wave and low frequency gate admittance measurements offer different perspectives onto the depletion of the 2D electron gas.  Figure 5 compares these two different techniques by displaying both the normalized SAW velocity shift $\Delta v/v$ at 856 MHz and the low frequency conductivity $\sigma_{xx}$ as functions of gate voltage $V_g$ and density at $T$ = 300 mK.  We stress that these data are obtained from the same sample during the same cooldown from room temperature\cite{cycling}.  In particular, while systematic errors may exist in the conversion of gate voltage to density, such errors do not impact the comparison of the velocity shift and conductivity at a given gate voltage.  It is clear from the figure that the SAW velocity shift begins to occur near $V_g = -1.07$ V where $N_s \approx 8 \times 10^9 \rm cm^{-2}$.  At this gate voltage the conductivity determined from the gate admittance measurements is roughly $\sigma_{xx}\approx e^2/h$.  This is surprising since $e^2/h$ is roughly 50 times larger than $\sigma_M$.  According to Eq. 1 the velocity shift at such a conductivity should be negligible ($\Delta v/v \approx 4 \times 10^{-4}$).  Equally surprising is the persistence of noticeable variation in $\Delta v/v$ down to $\sigma_{xx}\approx 10^{-5}e^2/h$, some 2000 times less than $\sigma_M$. These qualitative observations are rendered concrete by the dashed line in Fig. 5 which gives the normalized SAW velocity shift calculated by inserting the observed low frequency conductivity $\sigma_{xx}$ into Eq. 1.  This calculation predicts that the jump in $\Delta v/v$ should occur over a much narrower range of gate voltages and conductivities than is actually observed.  This fact is the central result of our work.

\begin{figure} [t]
\includegraphics[width=3.5in, bb=0 0 308 225]{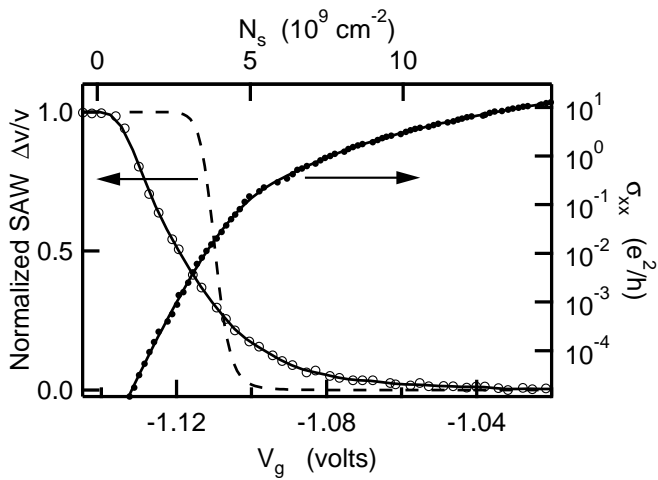}
 \caption{Low-frequency conductivity $\sigma_{xx}$ and SAW fractional velocity shift $\Delta v/v$ vs. gate voltage and 2D density\cite{cycling} at $T$ = 300 mK.  The dashed line is the prediction of Eq. 1 with the measured $\sigma_{xx}$ inserted.
 \label{fig5}}
\end{figure}
The conflict between the SAW velocity shift data and the low frequency conductivity measurements suggests that the 2D electron gas in our sample becomes inhomogeneous near depletion.  Obviously, in the presence of strong inhomogeneities Eq. 1 for the SAW velocity shift cannot be expected to apply.  For example, if an otherwise high conductivity ($\sigma_{xx}\gg \sigma_M$) 2DES is pierced by a number of holes, a SAW velocity shift can be expected.  In a gated device like ours, the magnitude of the shift will depend on the size of the holes relative to the distance $d$ to the metal gate\cite{gate}, their spatial distribution, and the fraction $f$ of the sample area still occupied by electrons.  Similarly, if the 2DES has become so depleted that no bulk conduction is possible and only isolated pockets of electrons remain, the SAW velocity shift will nonetheless register their presence and eventual depletion at more negative gate voltage.  These observations, which admittedly gloss over many subtleties, nonetheless offer a qualitative explanation for the basic facts displayed in Fig. 5.

All semiconductor-based 2D electron systems are inhomogeneous.  If the average density is high the inhomogeneities are unimportant.  Near depletion, however, inhomogeneities dominate the system behavior.  One obvious source of such inhomogeneities is statistical fluctuations in the Si donor population\cite{efros0,nixon}.  The 2DES, which is situated a distance $d_s$ from the Si donor sheet, responds to these flucutations by varying its own density.  Pikus and Efros\cite{efros} have shown that for randomly distributed donors with areal concentration $C$, the rms density fluctuation of the 2DES is $\delta N_s = \sqrt{C/8\pi d_s^2}$.  Assuming $C$ equals the 2DES density of the as-grown sample, $1.4 \times 10^{11} ~\rm cm^{-2}$, this formula gives $\delta N_s \approx 1 \times 10^{10} \rm~cm^{-2}$ for our sample in which $d_s = 70$ nm.  While this number seems even larger than needed to understand the results in Fig. 5, we remark that independent experiments\cite{jpecompress} have suggested that the Pikus-Efros formula overestimates $\delta N_s$ by about a factor of 2 in some samples, presumably because the donor distribution is not truly random.  If donor density fluctuations are the dominant source of 2DES density inhomogeneity in our sample, then the minimum length scale for these inhomogeneities will be the donor setback distance $d_s$.  This is qualitatively consistent with our observation that the shape of the SAW velocity shift near depletion is independent of frequency since our minumum SAW wavelength is $\lambda_{min}$ = 2.2 $\mu$m $\gg d_s$.

The preceding discussion suggests that the onset of insulating behavior in the 2DES in our sample is roughly coincident with the onset of inhomogeneity-induced SAW velocity shifts, at a density near $7 - 8 \times 10^9 \rm ~ cm^{-2}$.  At lower densities the system becomes increasingly inhomogeneous and more strongly insulating.  At the lowest gate voltages, where $\sigma_{xx} \ll \sigma_M$, the still changing velocity shift suggests a system of virtually isolated puddles of electrons with a tiny, very strongly temperature-dependent, hopping conductivity.  These findings are in good agreement with recent local compressibility measurements on similar GaAs-based 2D electron systems\cite{yacoby}.  Thus the apparent 2D metal-insulator transition is very likely strongly influenced by inhomogeneity and may be better described as a percolation transition than a homogeneous quantum phase transition. A comprehensive theory of SAW propagation in gated samples with inhomogeneous 2D electron systems near the percolation threshold is needed in order to pursue a more quantitative anaysis of this question.

In conclusion, we have studied the conductivity of a low density 2D electron gas via two different techniques, surface acoustic wave propagation and low-frequency gate admittance measurements.  We find that the two methods are in strong conflict at low density.  This conflict ensues near the apparent metal-insulator transition and progressively worsens as the 2D system is depleted.  We suggest that these results reflect the inhomogeneous nature of real 2D systems at very low density.  

We thank S.H. Simon and S. Das Sarma for enlightening discussions.  This work was supported by the DOE under Grant No. DE-FG03-99ER45766 and the NSF under Grant No. DMR-0242946.


\begin{thebibliography}{99}

\bibitem{krav}S.V. Kravchenko, G.V. Kravchenko, J.E. Furneaux, V.M. Pudalov, M.D'Iorio, Phys. Rev. B {\bf 50}, 8039 (1994).

\bibitem{abrahams}E. Abrahams, S.V. Kravchenko, and M. Sarachik, Rev. Mod. Phys. {\bf 73}, 251 (2001).

\bibitem{dassarma1}S. Das Sarma and E.H. Hwang, cond-mat/0411528.

\bibitem{efros0}A.L. Efros, Solid State Commun. {\bf 70}, 253 (1989).

\bibitem{nixon}J.A. Nixon and J.H. Davies, Phys. Rev. B {\bf 41}, 7929 (1990).

\bibitem{yacoby}S. Ilani, A. Yacoby, D. Mahalu, and Hadas Shtrikman, Phys. Rev. Lett. {\bf 84}, 3133 (2000).

\bibitem{meir}Y. Meir, Phys. Rev. Lett. {\bf 83}, 3506 (1999).

\bibitem{dassarma2}S. Das Sarma, {\it et al.}, Phys. Rev. Lett. 94, 136401 (2005).

\bibitem{Wixforth}A. Wixforth, {\it et al.}, Phys. Rev. B {\bf 40}, 7874 (1989).

\bibitem{Willett}R.L. Willett, {\it et al.}, Phys. Rev. Lett. {\bf 65}, 112 (1990).

\bibitem{simon}$\sigma_M$ has been calculated by S.H. Simon (Phys. Rev. B {\bf 54}, 13878 (1996)) for ungated samples; the value quoted here includes the screening effect of a gate.

\bibitem{lilly}M.P. Lilly, {\it et al.}, Phys. Rev. Lett. {\bf 82}, 394 (1999).

\bibitem{strain} The strain-induced linear portions of the velocity shift data at higher gate voltage are, as expected, frequency independent.

\bibitem{calib} Minima in Im($Y$) correspond to the filling of an integer number of Landau levels and thus give an unambiguous determination of $N_s$ at a discrete set of $V_g$ values.  At the lowest densities the linear relationship of $N_s$ and $V_g$ must be extrapolated and the possibility of small systematic error arises. See Ref.\cite{jpecompress}.

\bibitem{cycling} The slight ($\sim$10 mV) gate voltage shift in the conductivity data between Figs. 4 and 5 are representative of the roughly 1\% variations in the ungated 2DES density we observe from one cool-down to the next.

\bibitem{gate} The presence of the gate cuts off the ability of the the electron gas to screen the SAW electric field beyond distances of order $d = 0.6~\mu$m, which is substantially less than the smallest SAW wavelength employed here, $\lambda_{min} = 2.2~\mu$m.

\bibitem{efros}F.G. Pikus and A.L. Efros, Phys. Rev. B {\bf 47}, 16395 (1993).

\bibitem{jpecompress} J.P. Eisenstein, L.N. Pfeiffer and K.W. West, Phys. Rev. B {\bf 50}, 1760 (1994).

\end{thebibliography}
\end{document}